\documentclass{llncs}
\usepackage{etoolbox}
\newtoggle{disc}
\toggletrue{disc}

		\usepackage{tikz} % I am using tikz 2.??

\begin{document}

\title{Implicit Consensus: Blockchain with Unbounded Throughput}
\titlerunning{Implicit Consensus}  % abbreviated title (for running head)
\author{Zhijie Ren\inst{1} \and Kelong Cong\inst{2} \and
Johan Pouwelse\inst{2} \and Zekeriya Erkin\inst{1}}
\authorrunning{Zhijie Ren et al.} % abbreviated author list (for running head)
\tocauthor{Zhijie Ren, Kelong Cong, Johan Pouwelse, Zekeriya Erkin}
\institute{Cyber Security Group, Dept. of Intelligent Systems\\
\and
Distributed Systems Group,  Dept. of Software Engineering \\
Delft University of Technology, Mekelweg 4, Delft, the Netherlands\\
\email{z.ren@tudelft.nl},}

\maketitle              % typeset the title of the contribution

\begin{abstract}
Recently, the blockchain technique was put in the spotlight as it introduced a systematic approach for multiple parties to reach consensus without needing trust. However, the application of this technique in practice is severely restricted due to its limitations in throughput. In this paper, we propose a novel consensus model, namely the implicit consensus, with a distinctive blockchain-based distributed ledger in which each node holds its individual blockchain. In our system, the consensus is not on the transactions, but on a special type of blocks called Check Points that are used to validate individual transactions. Our system exploits the ideas of self-interest and spontaneous sharding and achieves unbounded throughput with the transaction reliability that equivalent to traditional Byzantine fault tolerance schemes.
\end{abstract}
\section{Introduction}\label{s:intro}

Blockchain, introduced and firstly applied in Bitcoin \cite{nakamoto}, is one of the hottest techniques in the information technology at this moment. Researchers see a huge potential in this technique and believe that it can be used as a systematic approach to replace trust. More precisely, in a network in which nodes do not trust each other, blockchain provides a way for the nodes to reach consensus and cooperate without a third party or a central authority. Typically, a blockchain technique is a distributed append-only database which consists of two components. First, as its name suggests, the database is an ordered sequences of blocks which are chained together. The newly generated data forms a new block and chains to the existing chain with a digest of the cryptographic hash function of the previous block. Second, a consensus algorithm is used for the network to agree on the new block that will be appended to the chain.

\subsection{Problem Statement}\label{ss:ps}

One of the major problems of the blockchain technique is the scalability, which is caused by the limitation of the current consensus algorithms. Reaching consensus for a number of nodes that might be malicious, commonly known as the Byzantine fault tolerance (BFT) problem \cite{lamport}, has been extensively studied for over 30 years. The problem could be described as the following. In an asynchronous network with malicious nodes, a BFT scheme should have the following properties.
\begin{itemize}
\item {\bf Agreement:} If an honest node propose a vector $v$, then all honest nodes agree with $v$.
\item {\bf Correctness:} If an honest node decides $v$, then $v$ must be proposed by at least one honest node.
\item {\bf Termination:} If an honest node propose $v$, then all honest node will eventually decide a vector.
\end{itemize}
In general, these properties are not feasible for a full-asynchronous network \cite{flp}. However, with some assumptions like weak synchronous or probabilistic agreement, these properties can be achieved with message complexity of $O(N^2)$ \cite{bracha,castro} when the number of the adversary is less than $\lfloor \frac{N-1}{3} \rfloor $.
%Bitcoin and its proof-of-work (POW) scheme opened up a new horizon on this problem.
%, in which the consensus is not final, but probabilistic. Moreover, 
%In Bitcoin, incentives are introduced to put cost on malicious behaviors, which limits the capability of the adversaries. It thus allows reaching consensus in a network with thousands of nodes, as long as the computational power of the malicious nodes is less than 1/4 of the total computational capability \cite{majority}. 

%However, POW based public blockchains like Bitcoin and Ethereum \cite{ethereum} have rather low throughputs, e.g., Bitcoin can have at most seven transactions per second. This throughput is incomparable to the throughput of thousands of transactions per second achieved by traditional BFT algorithms \cite{croman}. Currently, there does not exist a blockchain system which achieves high throughput in a network with thousands of nodes due to the limitation in the consensus algorithm.

\subsection{State-of-the-Art}

{\bf Towards the scalability of BFT.} Many schemes have been proposed to increase the scalability of the traditional BFT schemes. Luu et al. divides the network into subgroups uniformly and runs BFT within the subgroup \cite{elastico}. Other schemes like \cite{700bft,byzcoin} opportunistically apply simpler schemes for good network scenarios, usually with message complexity of $O(N)$, and use ``safe'' schemes like PBFT \cite{castro} as backup. In networks with good connectivity and limited amount of malicious node, these approaches result in message complexity of $O(N)$ but risks even higher latency comparing to traditional BFT schemes in bad situations.

{\bf Increasing the throughput of POW.} By introducing mining costs for messaging and incentives to compensate the costs, Bitcoin proposed a scalable solution for BFT that reduces the message complexity to $O(N)$ and functions in large networks with thousands of nodes. However, this scheme requires synchrony amongst the majority, which restricts the throughput of Bitcoin since the block size and frequency are then limited \cite{croman}. Scalable Bitcoin schemes like Bitcoin-NG and lightning network \cite{ng,ln} separate the consensus from the contents of the block. As a result, the synchrony is not restricted by the block size and frequency. However, extra economical punishment methods are required since the costs and incentives of the traditional POW no longer gives protection to the validity of the contents.

{\bf Leader selection.} The schemes of \cite{nakamoto,ng,ln} can be seen as two parts: an economical rewarded leader election process and an economical guaranteed validation process. Schemes like Byzcoin, ALGORAND, hybrid consensus, and the sleepy model of consensus \cite{byzcoin,algorand,hybrid,sleepy} keep the leader selection part. Then, instead of a sole leader, they select a group of leaders and the validation is guaranteed by BFT algorithms. This approach has some advantages over both POW and BFT. However, extra effort is needed to guarantee that the leader selection is unbiased, i.e., the malicious nodes cannot predict or control the selection results.

{\bf Sharding.} The sharding idea proposed in Etheruem and Omniledger \cite{ethshard,omni} has received a lot of attention recently. The basic idea of sharding is that the network is divided into shards and the intra-shard transactions are only agreed, validated, and recorded in the nodes of that shard. Then, some nodes are selected to validate and record the inter-shard transactions. In networks where the transactions pattern is rather isolated, sharding has a potential of achieving unbounded performance, i.e., less than $O(N)$ message complexity. The reason is that the transactions are not mandatory to be broadcast to the whole network. However, the choice of the shards for a network to achieve optimal performance remains a challenge.

\subsection{Implicit Consensus}\label{ss:ic}

In this paper, a novel consensus model, namely the {\em implicit consensus}, is proposed, which achieves unbounded throughput for a distributed ledger type of blockchain system with equivalent reliability on validated transactions to traditional BFT based schemes. It has some similarities to some of the existing ideas like the side-chain ideas used in \cite{ln} and sharding \cite{ethshard,omni}. However, compare to the existing schemes, the main innovations of our consensus model are the following.

\begin{itemize}
\item Replacing the {\em termination} property of BFT schemes by the {\em self-interest phenomenon}. In other words, for each transaction, our scheme does not guarantee consensus. However, because it is in the interest of the related parties, e.g., the issuer of this transactions, to convince the other nodes that this transaction is valid. Thus, nodes are encouraged to prove the validity and agreement of their transactions to as many nodes as they can and we guarantee that the malicious nodes cannot prevent them from doing that. It is arguably more close to real-life scenario. To distinguish this from traditional consensus in which each transaction has explicitly reached consensus (termination property), we call this {\em implicit consensus}. With implicit consensus, our scheme is scalable since the message complexity is reduced to $O(N)$.
\item Spontaneous sharding. For each transaction, we prove that the other two properties of BFT, {\em agreement and correctness}, will hold as long as the transactions is locally validated by our validation scheme. This is done with collecting some faction of the total transactions of the network related to this transaction, called {\em proofs}. As a result, the network is spontaneously sharded since rational nodes will optimize their storage and transmission costs by only validating and recording the proofs of their transactions. There is no need for mechanisms that allocate the transaction set needed to be validated and recored for each node. Similar as sharding, our scheme achieves unbounded performance, i.e., the message complexity is less than $O(N)$ since transactions are not mandatory to be broadcast to the whole network.
\item Uncompromised reliability. Although our scheme has its similarity to side-chain schemes like lightning network \cite{ln}, we guarantee that the validated transactions on the the individual chains (can be seen as side-chains) are as reliable as they are on the main chain. This is fundamentally different from side-chain approaches like lightning network, in which the reliability is not guaranteed by the main chain, but by the locked deposit.
\end{itemize}

\iftoggle{disc}{}{In this paper, the implicit consensus is achieved by introducing some special blocks called Check Points (CPs) and using a BFT algorithm on these CPs, instead of the individual transactions, to reach consensus. Such a scheme achieves an unbounded throughput, since the consensus on individual transaction is no longer necessary and transaction throughput does not depend on the BFT algorithm. Then, we let the nodes to validate the transaction of the others or prove their own transactions by their own. The validation result is proved to be correct and consistent.}

\subsection{Structure of Our System}

We consider an asynchronous network with $N$ nodes and $f \leq \lfloor \frac{N-1}{3} \rfloor$ adversaries. To achieve the implicit consensus, we propose a permissioned blockchain-based distributed ledger consisting of four layers: transactions, individual blockchains, the consensus scheme, and the validation scheme. The first layer of our system is {\bf transactions}, which are defined in a similar fashion as Bitcoin.
The second layer is {\bf individual blockchains}. In our system each node has its own genesis block and blockchain, in which only transactions that related to the node itself are recorded. Besides the blocks that consists of transactions which are called {\em Transaction Blocks (TBs)}, a special type of blocks called {\em Check Point (CP)} is introduced. The CPs contain no transaction, but some already established consensus and a hash of the previous block. The third layer of our scheme is the {\bf consensus scheme}, in which we plug in one of the existing Byzantine fault tolerance (BFT) schemes like \cite{castro,700bft,scp,miller} to reach consensus on the hashes of the CPs. One of the fundamental differences between our system and other blockchain systems is that the consensus is reached only on the hashes of the CPs instead of all transactions. As a result, if some CP reached consensus, the transactions that came before the CP are tamper-proof. However, this tells nothing about the validity of the transactions in these parts. Hence, in the fourth part, a {\bf validation scheme} is used to validate individual transactions. The validation scheme is executed locally and only based on point-to-point communications. Since the CPs in the consensus have ``sealed'' the chains, the authenticity and integrity of the chains can be easily verified. We prove that although the validation scheme is run locally, the result is correct and consistent for all honest nodes, which suggests the agreement and correctness properties, i.e., implicit consensus.% This scheme enables an unbounded throughput, since the duration of making transactions is no longer limited by the performance of the consensus algorithm. Practically it is still bounded by the running time of the validation process. However, that is much faster compares to traditional BFT algorithms and proof-of-work (POW) type of consensus schemes since it only requires point-to-point communications. Also, the validation scheme can be executed by individual nodes in parallel independently.

% \begin{figure}
% \centering

% \begin{tikzpicture}[scale=.35]
% \tikzstyle{->}=[-latex];

% \node[fill=white,inner sep=1pt] at (0,1) {Transactions};
% \node[fill=white,inner sep=1pt] at (0,5) {Individual Blockchains};
% \node[fill=white,inner sep=1pt] at (0,9) {Consensus Scheme};
% \node[fill=white,inner sep=1pt] at (0,13) {Validation Scheme};

% \draw (-8,0) rectangle (8,2);
% \draw (-8,4) rectangle (8,6);
% \draw (-8,8) rectangle (8,10);
% \draw (-8,12) rectangle (8,14);

% \draw[->, line width=3pt] (0,2) -- (0,4);
% \draw[->, line width=3pt] (0,6) -- (0,8);
% \draw[->, line width=3pt] (0,10) -- (0,12);

% \end{tikzpicture}
% \caption{Conceptual Structure of the Implicit Consensus System}\label{fig:str}
% \end{figure}

\subsection{Content of the Paper}

Firstly, we introduce the four-layer system in Section~\ref{s:nut}. Then we show the necessary theorems and proofs in Section~\ref{ss:proofs}. The performance of our system is discussed with the focus on throughput and reliability in Section~\ref{s:adv}. In Section~\ref{s:fw}, we conclude our work.

\section{Our System}\label{s:nut}

%In this section, we introduce our four-layer system which achieves the implicit consensus.% Sometimes terms are used with definition given in the later parts of this paper. In this case, we present the terms associated with (sub)section numbers in brackets to guide readers to the formal definitions of those terms.

\subsection{Transactions}\label{ss:trans}

We consider a transaction based value-exchange blockchain system similar to Bitcoin, i.e., a distributed ledger. The $s$-th transaction from node $i$ to node $j$ is denoted by $tr(i \to j, s)$. In this paper, a cryptographic hash function is denoted by $Y=H(X)$, in which $Y$ is called the {\em digest} of $X$. Furthermore, we assume that there is a public-key infrastructure (PKI) such that each node holds a secret private key while the corresponding public key is known to all other nodes. A transaction $tr(i \to j, s)$ contains the following information.
\begin{itemize}
\item The sender and the receiver, i.e., $i$ and $j$.
\item A unique serial number $s$ one-to-one mapped to this transaction $tr(i \to j,s)$.
\item The indices of the sources of this transactions (see Definition~\ref{def:ti}). The sources are denoted by ${\cal S}(tr(i \to j, s))$, which is a set of previous transactions send to $i$. This is equivalent to the input of Bitcoin.
\item The value of this transaction $V_t$ and the remaining value $V_r$ after this transaction. This is equivalent to the output of Bitcoin.
\item A digital signature created by node $i$, which is the digest of the aforementioned items encrypted with the private key of $i$.
\end{itemize}
Here, we only consider two-party transactions. The transactions are only recorded in the chains of the related node. Hence, a transaction $tr(i \to j, s)$ is recorded in the chains of nodes $i$ and $j$.

\subsection{Individual Blockchains}\label{ss:chain}

We consider a permissioned network with $N$ nodes and each node has its own blockchain. The blockchains are denoted by ${\cal B}_i, i \in \{1,2,\ldots, N\}$ \footnote{This definition is slightly naive since malicious nodes could have multiple versions of their blockchains. For the sake of easier comprehension, we use this definition here and will further address this problem in Section~\ref{ss:proofs}.}. A blockchain ${\cal B}_i$ is then defined as an ordered set of blocks $\{B_i(1), B_i(2), \ldots\}$, in which each block contains a digest of its previous block, i.e., block $B_i(j), j>1$ will contain $H(B_i(j-1))$. The genesis blocks (the first blocks in the chains) $B_i(1)$ are distinctive and contain the information about their unique identities and the initial balance of each node\footnote{Our system only focus on the reliable value exchange, thus we assume that there exists some pre-established agreement on the initial balance for the nodes.}. The initial balance can be seen as a transaction without sources and has an unique index (see Definition~\ref{def:ti}). Furthermore, this transaction is valid if the corresponding genesis block is included in some consensus. Otherwise, it is invalid (see Subsections~\ref{ss:cons} and \ref{ss:val}).

There are two types of blocks in the chains: transaction blocks (TBs) and check points (CPs).  We assume all genesis blocks are CPs. TBs are used to record the transactions and CPs are used for the consensus scheme. Now we introduce these two types of blocks.

\subsubsection{Transaction Blocks}

As its name suggests, TB is used to record the transactions. We denote the $k$-th TB in ${\cal B}_i$ by $T_i(k)$. Then, if $T_i(k)$ is the $j$-th block in ${\cal B}_i$, we say that $T_i(k) \equiv B_i(j)$. A transaction block $T_i(k)$ consists of a digest of the previous block and $M$ transaction messages $t_i(k,m)$, i.e., if $T_i(k) \equiv B_i(j)$, then $T_i(k)$ consists of $[H(B_i(j-1)), t_i(k,1), t_i(k,2), \ldots, t_i(k,M)]$.

\begin{definition}[Transaction Message]
A {\bf transaction message} $t_i(k,m)$ is the $m$-th message in the $k$-th transaction block. It consists a transaction $tr(a \to b,s)$ where $a=i$ or $b=i$.
\end{definition}

\begin{definition}[Transaction Index]\label{def:ti}
If a transaction $tr(a \to b,s)$ is in the transaction message $t_i(k,m)$, then a vector of $[i,k,m]$ are called the {\bf index} of this transaction.
\end{definition}

Note that since a transaction is written in the chains of both the sender and the receiver, a valid transaction should have two indices. Also, the two transaction messages of a transaction are identical.

\subsubsection{Check Points}

CPs are a special type of blocks which contain no transaction. Instead, they contain some established consensus. We use the consensus scheme to reach consensus on the digests of the CPs of this round. Similar to the TBs, we denote the $k$-th CP in ${\cal B}_i$ by $C_i(k)$. Then, if $C_i(k)$ is the $j$-th block in ${\cal B}_i$, we say that $C_i(k) \equiv B_i(j)$. The CPs are defined as follows.

\begin{definition}[Check Point (CP)]
A {\bf check point} consists of a digest of the previous block and the consensus established in the previous round (see Subsection~\ref{ss:cons}).% denoted by $CON(r-1), r\in \{1,2,\ldots\}$, i.e., assume that $C_i(k) \equiv B_i(j)$, then $C_i(k)$ consists of $H(B_i(j-1))$ and $CON(r-1)$.
\end{definition}

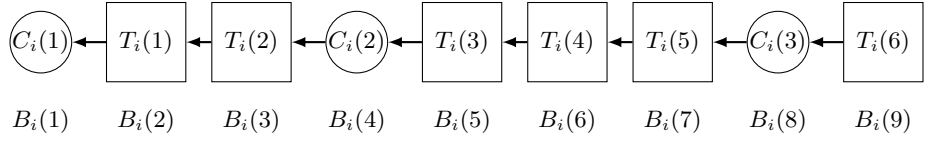
\begin{figure}[t!]
\centering
\tikzstyle{TB}=[minimum size=30pt, inner sep=0pt, draw]
\tikzstyle{CP}=[circle,minimum size=15pt, inner sep=0pt, draw]
\tikzstyle{edge}=[-latex, black, thick]
%\tikzstyle{dot}=[edge, dotted]
%\tikzstyle{dash}=[edge, densely dashed]
%\tikzstyle{loosedo}=[edge, loosely dotted]
%
%
\begin{tikzpicture}[scale=0.7]
\node[CP] (b1) at (0,0) {$C_i(1)$};
\node[TB] (b2) at (2,0) {$T_i(1)$};
\node[TB] (b3) at (4,0) {$T_i(2)$};
\node[CP] (b4) at (6,0) {$C_i(2)$};
\node[TB] (b5) at (8,0) {$T_i(3)$};
\node[TB] (b6) at (10,0) {$T_i(4)$};
\node[TB] (b7) at (12,0) {$T_i(5)$};
\node[CP] (b8) at (14,0) {$C_i(3)$};
\node[TB] (b9) at (16,0) {$T_i(6)$};

\node (bb1) at (0, -1.5) {$B_i(1)$};
\node (bb2) at (2, -1.5) {$B_i(2)$};
\node (bb3) at (4, -1.5) {$B_i(3)$};
\node (bb4) at (6, -1.5) {$B_i(4)$};
\node (bb5) at (8, -1.5) {$B_i(5)$};
\node (bb6) at (10, -1.5) {$B_i(6)$};
\node (bb7) at (12, -1.5) {$B_i(7)$};
\node (bb8) at (14, -1.5) {$B_i(8)$};
\node (bb9) at (16, -1.5) {$B_i(9)$};

\draw[edge] (b2) -- (b1);
\draw[edge] (b3) -- (b2);
\draw[edge] (b4) -- (b3);
\draw[edge] (b5) -- (b4);
\draw[edge] (b6) -- (b5);
\draw[edge] (b7) -- (b6);
\draw[edge] (b8) -- (b7);
\draw[edge] (b9) -- (b8);

\end{tikzpicture}
\caption{Example of a blockchain of node $i$ with six TBs and three CPs.}\label{fig:cp}
%\vspace{-2mm}
\end{figure}

The relationship between blocks, TBs, and CPs is shown in Figure~\ref{fig:cp}.

\subsection{Consensus Scheme}\label{ss:cons}

Our consensus scheme is a {\em consensus process} used repetitively in {\em rounds}. In each round, the consensus process is used to reach consensus on consensus messages (CMs), which is defined as follows.

\begin{definition}[Consensus Message (CM)]
A {\bf consensus message} of node $i$ in round $r$ denoted by $M_i(r)$ consists of the following information.
\begin{itemize}
\item $i$ and $r$.
\item The digest of a CP $C_i(k)$ which has not been {\bf included} in any consensus. We call a CP is {\bf included} in some consensus if and only if the digest of this CP is in a CM and that CM has reached consensus.
\item The position of the CP $C_i(k)$ and its previous CP $C_i(k-1)$ in the chain, i.e., two numbers $j$ and $j'$ that $C_i(k) \equiv B_i(j)$ and $C_i(k-1) \equiv B_i(j')$.
\item A digital signature of $i$, which is the digest of the aforementioned items encrypted by the private key of $i$.
\end{itemize}
\end{definition}

The consensus process of round $r$ starts when the consensus process of round $r-1$ is complete and the consensus result, denoted by $CON(r-1)$, is acknowledged by all honest nodes. Now, we describe the steps of the consensus process for node $i$ of the $r$-th round.
\begin{itemize}
\item {\bf Step 1:} After $CON(r-1)$ is obtained, if a CP from node $i$ is included in $CON(r-1)$, it generates a new CP with $CON(r-1)$ and appends it to its chain.
%\item If no new CP is generated, node $i$ waits for time $\tau_{i,0}$ and broadcasts the digest of its latest CP that does not reach consensus and its previous CP that reached consensus in $CON(r')$. Furthermore, the indexes of the node, the two CPs, and a digital signature is associated with this message, which is called a consensus message. We denote this message by $M'=[H(B_i(j')), H(B_i(k')),i, j',k',D']$.
%\item If a new CP $B_i(k)$ is generated, it waits for time $\tau_{i}$ and broadcast the digest of this CP and its previous CP that reaches consensus in $CON(r-1)$. Furthermore, a digital signature is associated with this message. Let us denote this message by $M=[H(B_i(j)), H(B_i(k)),i,j,k,D]$

%\item If no new CP is generated, node $i$ waits $\tau_{i,0}$ and broadcast its latest CP that does not reach consensus and its previous CP that reached consensus in $CON(r')$. Furthermore, the indexes of the node, the two CP, and a digital signature is associated with this message, which is called a consensus message. We denote this message by $M'=[C_i(j'), C_i(k'),i, j',k',D']$.
%\item If a new CP $C_i(k)$ is generated, it waits $\tau_{i}$ and broadcast this CP and its previous CP that reaches consensus in $CON(r-1)$. Furthermore, a digital signature is associated with this message. Let us denote this message by $M=[C_i(j), C_i(k),i,j,k,D]$.
\item {\bf Step 2:} It generates a new CM using its latest CP, and uses this as its input for the consensus process of this round.
\item {\bf Step 3:} A BFT algorithm is used to reach agreement on a set of input CMs of this round. The following CMs will be excluded from this consensus process by all honest nodes:
\begin{itemize}
\item The CMs of incorrect rounds.
\item The CMs with incorrect digital signatures.
%\item The messages with incorrect information.
\item The CP included in the CM has already been included in some previous consensus.
\item The index of the previous CP that has been used by a CM which has already reached consensus. Also known as a ``fork''.
\end{itemize}
%Any asynchronous BFT scheme can be plugged in here for the consensus process.
\item {\bf Step 4:} Output the result of this consensus process denoted by $CON(r)$, which is a vector consisting the CMs ordered by $i$.
\end{itemize}

For {\bf Step 3}, any consensus algorithm that satisfies the agreement, termination, and correctness properties introduced in Subsection~\ref{ss:ps} can be used.
% \begin{itemize}
% \item {\bf Agreement:} If an honest node sends a CM, then all honest nodes agree with this CM in the consensus.
% \item {\bf Termination:} If all honest nodes send their CMs, then they reach some consensus eventually.
% \item {\bf Correctness:} If a CM is in the consensus, then it must have been send by some node.
% \end{itemize}
%In this paper, we use the Honey-badger BFT (HBFT) since it meets these conditions and has a good performance. 
Some of the choices are \cite{700bft,scp,miller}, which tolerant less than $\lfloor \frac{N-1}{3} \rfloor$ malicious nodes. Here, the {\em honest} and {\em malicious} nodes are defined as the following.

\begin{definition}[Honest Node\footnote{Our definition of the honest node is stronger than that in some other literature, in which the honesty is round based, i.e., a node is considered honest if it does not conduct malicious behaviors in that round. However, this strong assumption is solely because that we would like to keep the core system as simple as possible. We will later show in Subsection~\ref{ss:reli} that the definition can be easily weaken to the round based honesty by simple mechanisms.}]\label{def:honest}
An honest node is a node that creates correct CM messages and cooperates in the consensus process to reach consensus. Moreover, it always validates all of its transaction and only make transactions with correct information, sufficient balance, and validated sources which have not been used in previous transactions (see Subsection~\ref{ss:val}).
\end{definition}

\begin{definition}[Malicious Node]
A malicious node can do anything to prevent consensus, creates any kind of transaction, and manipulates its chain, e.g., creates forks, to confuse honest nodes. Moreover, malicious nodes can collude. However, we assume that they cannot break the hash function or the asymmetric encryption.
\end{definition}

The consensus result $CON(r)$ is a vector consisting all the CMs that have reached consensus in this round. We denote the already established consensus till round $r$ by ${\cal CON}(r)= \{CON(1), CON(2), \ldots, CON(r)\}$. By the properties of the BFT algorithm, $CON(r)$ is known and should be recorded in the blockchains of all honest nodes by the end of round $r$.

A CP included in ${\cal CON}(r)$ guarantees the tamper-proof property in the sense that the transactions previous to this CP are unforgeable. Here, we introduce the term {\em correct piece}.
\begin{definition}[Correct Piece]\label{def:cp}
An ordered set of blocks $\{B_i(j),\ldots, B_i(k)\}, k-j \geq 1$ is called a {\bf piece} if $B_i(j) \equiv  C_i(\ell) , B_i(k)\equiv C_i(\ell+1) $. This piece is {\bf correct} if and only if:
\begin{itemize}
\item $C_i(\ell)$ and $C_i(\ell+1)$ are both included in ${\cal CON}(r)$.
\item All digests in $\{B_i(j),B_i(j+1),\ldots, B_i(k)\}$ are correct.
\end{itemize}
\end{definition}

%Here, we describe our consensus process in detail. The details of HBFT is omitted for space limitation, for which we refer to \cite{miller}.

\subsection{Validation Scheme}\label{ss:val}

With the established consensus ${\cal CON}(r)$, the honest nodes can validate individual transactions without any knowledge of the transaction in advance and thus achieves the implicit consensus. In this subsection, we introduce our validation scheme. Note that there are two fundamental difference between our system and other blockchain systems. First, invalid transactions are allowed in our blockchains. Second, there is no globally agreed blockchain and each node might have different observations of the blockchains of the network. Hence, we will first give the definition of the valid transactions and invalid transactions in our system. Then, we show that our validation scheme allows the honest nodes to check the validity of the transactions.

\subsubsection{Validity and Conditions for Validation}

In general, the validity of transactions should be a global and unambiguous property that is independent of the observation of the network by any specific node. In a ledger, a valid transaction should have correct format, sufficient balance, unspent sources, and be signed by the private key of the sender. Besides, each system has its own definition in the validity of the transactions, e.g., a valid transaction in Bitcoin should be in the longest chain for a sufficient long period of time.
In our system, a valid transaction should have two messages in both chains of the senders and the receivers. Furthermore, it should also be included in the authentic chain. Hence, we have the following definition.

\begin{definition}[Validity]\label{def:vali}
The validity conditions of a transaction $tr(i \to j, s)$ are the following.
\begin{itemize}
\item {\bf Two Messages:} The transaction is written in exact two identical messages included in the chains of both sender and receiver, respectively.
\item {\bf Correct Chains:} The two messages are in correct pieces (Definition~\ref{def:cp}) and all pieces that are previous to these two pieces in the corresponding chains are also correct.
\item {\bf Correct Messages:} These messages are correct in the sense that all information are correct and signed with the private key of $i$ 
%\item {\bf Authorization:} The transaction is made by the authorized node, which is the node that has the private key of the sender.
\item {\bf Valid Sources:} The transactions which used as source are valid.
\item {\bf No Double Spending:} The sources have not been used by other transactions.
\item {\bf Sufficient Balance:} The transaction value $V_t$ plus the remaining value $V_r$ equals to the sum of all the remaining values of all sources. 
\end{itemize}
A transaction is {\bf valid} if it satisfies all the validity conditions in the observation of any node in the network. Otherwise, it is an {\bf invalid} transaction.
\end{definition}

Then, to achieve implicit consensus, a validation scheme should satisfy the following conditions.
\begin{itemize}
\item {\bf Liveness:} All transactions can be verified by the validation scheme eventually, the result is either ``validated''(verified as valid) or ``falsificated'' (detect as fraud).
\item {\bf Correctness:} All transactions validated by the honest nodes are valid. All transactions falsificated by the honest nodes are invalid.
% \item {\bf Correctness:} Valid transactions can be validated by honest nodes. Invalid transactions cannot be validated by honest nodes.
% %\item {\bf Unambiguousness:} A transaction cannot be both validated and falsificated.
% \item {\bf Liveness:} A valid transaction can be validated before some time $t$.
% %\item {\bf Unambiguousness:} If a transaction is validated (falsificated) by an honest node at a certain time $t$, it will also be validated (falsificated) by that node at another time $t'>t$ if it can be validated (falsificated).
% %\item {\bf Correctness:} A transaction that is locally validated (falsificated) by an honest node is also globally valid (fraudful).
%\item {\bf Consistency:} A validated transaction by an honest node in time $t$ will also be validated by another honest node in time $t' \geq t$ if it can be validated.
\end{itemize}
Clearly, if a transaction satisfies the above conditions, it implies that the agreement and correctness properties are satisfied with our validation scheme, i.e., they are in implicit consensus.

% {\em Weak Liveness}.
% \begin{itemize}
% \item {\bf Weak Liveness:} If a transaction has been validated (falsificated) by an honest node at a certain time $t$, it can be validated (falsificated) by an honest node at another time $t'>t$.
% \end{itemize}

% Moreover, we also introduce the idea of falsificate to our validation scheme, since finding fraud transactions is also very important in our system. One fundamental difference between our system and traditional blockchain systems is that the {\em invalid transactions} is allowed to appear in the blockchains. Hence, it is crucial to tell the difference between a fraud transaction, which can never be validated, and an uncompleted transaction, which might still be validated in the future. 

Our validation scheme consists of two parts: proof collection and validation process. Now we introduce our validation scheme by considering the case that node $u$ want to validate the transaction $tr(i \to j, s)$, denoted by a function $V_u(tr(i \to j,s))$.

\subsubsection{Proof Collection}

The proof collection is a process that a node requests all necessary information that it needs to validate a transaction, which is called the {\em proofs} of this transaction.
\begin{definition}[Proofs of a transaction]
The proofs of a transaction $tr(i \to j, s)$ consists of the following.
\begin{itemize}
\item All pieces of ${\cal B}_i$ from the first piece in the chain to the first piece which contains $tr(i \to j, s)$.
\item All pieces of ${\cal B}_j$ from the first piece in the chain to the first piece which contains $tr(i \to j, s)$.
\item For each source transaction of $tr(i \to j, s)$ or recursively the source of the sources until the initial balance in the genesis block, denoted by $tr(k \to l, s')$, all pieces of ${\cal B}_k$ signed by $k$ and ${\cal B}_l$ signed by $l$ from the first pieces to the ones containing $tr(k \to l, s')$.
\end{itemize}
The proofs of a transaction are {\em complete} if all the aforementioned items are collected. The proofs of a transaction are called {\em correct} if all the collected pieces are correct.
\end{definition}

To collect the proofs, three steps are taken by node $u$. All collected pieces are verified and the incorrect pieces are immediately discarded. Once the complete and correct proofs of the transaction are collected, the node terminates the proof collection and enters the validation process. If the complete proofs cannot be obtained within a certain time period, the transaction will be marked as ``undecided''. An undecided transaction could be validated in the future.

\begin{itemize}

\item {\bf Step 1:} It requests the transaction indices of $tr(i \to j, s)$ from either node $i$ or $j$.

\item {\bf Step 2:} It requests all the missing proofs from either node $i$ or $j$.

\item {\bf Step 3:} It broadcasts the request of the missing proofs to the whole network.

\end{itemize}

All the nodes are required to keeps the proofs of all transactions related to themselves.
%The node are encouraged to keep the proofs of other transactions to increase the liveness of the validation protocol (will be specified later).

%\subsection{Completion Check}

%Before the validation process, an {\em completion check} is done to check whether a transaction can be validated. The process of the completion check is given in the following definition.

%\begin{definition}[Completion Check]
%A transaction $tr(i \to j, s)$ fails the completion check and cannot be validated if the proofs of $tr(i \to j, s)$ are incomplete.
%Such transactions are called {\em undecided transactions}.
%\end{definition}

%Once a transaction passes the completion check, it enters the validation process.

\subsubsection{Validation Process}

%The validation process of a transaction is given in the following definition.

\begin{definition}[Validation Process for a Transaction]\label{def:vp}
A validation process of a transaction $tr(i \to j, s)$ includes the verification of the following items.
\begin{enumerate}
\item {\bf Two Messages:} The transaction with the serial number $s$ has two and only two identical messages $t_i(m,k)$ and $t_j(n,\ell)$.
\item {\bf Correct Messages:} All information in the messages is correct and signed with the private key of $i$. \label{stp:corm}
%\item The pieces of blockchains containing $t_i(m,k)$ and $t_j(m',k')$ are agreed.
%\item There exists $C_i(n)$ and $C_j(n')$ such that $T_i(k) \gets C_i(n)$ and $T_j(k') \gets C_j(n')$ and $C_i(n)$ and $C_j(n')$ are included in one or two $CON(r)$.
%\item Assume the serial numbers of the CP in the messages are $a$ and $b$. Then, $n\leq a+e$ and $n' \geq b+e$.
%\item The hash results in blockchains ${\cal B}_i$ and ${\cal B}_j$ from the first block till the blocks $B_i(n)$ and $B_j(n')$, respectively, are correct.
\item {\bf No Double Spending:} There are no forks for this transaction, i.e., there does not exist a validated transaction $tr'$ written in message $t_i(m',k')$ with $(k'=k, m' < m)$ or $k'<k$ and the source transactions ${\cal S}(tr') \cap {\cal S}( tr(i \to j, s)) \ne \emptyset$. \label{stp:fork}
\item {\bf Validated Sources:} All the source transactions of $tr(i \to j, s)$ are validated.
\item {\bf Sufficient Balance:} The transaction amount plus the remaining amount equals to the sum of the remaining amounts of all sources. All the amounts here are non-negative. 
\end{enumerate}
A transaction that passes or failed the validation process is called a {\bf validated transaction} or a {\bf falsificated transaction}, respectively.
\end{definition}

\section{Correctness of the System}\label{ss:proofs}

The correctness of our system is proved if the agreement, termination, and correctness conditions in Subsection~\ref{ss:cons} are satisfied for the consensus scheme and the liveness and correctness conditions in Subsection~\ref{ss:val} are satisfied for the validation scheme. The consensus conditions are guaranteed by the BFT schemes. For proofs we refers to the original papers of these schemes \cite{bracha,castro,miller}. Here, we prove the validation scheme satisfies the conditions of correctness and liveness. For the sake of space, all the proofs are included in the appendices.

%In this paper, we assume that the digests are collision free and the digital signatures are unforgeable. 

\subsection{Liveness}\label{ss:live}
The liveness condition is crucial since in our system, a transaction is only authentic when it is validated. 
However, as can be observed from our validation scheme, the liveness condition is in general not feasible since we allow the transaction to be ``undecided''.
Now, we give the following theorem and argue that our system is already reliable if we guarantee that the liveness condition holds for all transactions made by honest nodes.

\begin{theorem}[Liveness of the Honest Nodes]\label{th:pl}
If $i$ and $j$ are both honest nodes, the outcome of the validation scheme for a transaction $tr(i \to j,s)$ should be either validated or falsificated before time $t$.\footnote{We show that they will only be validated in Theorem~\ref{cor:proof}.}
\end{theorem}

Note that Theorem~\ref{th:pl} does not guarantee that all transactions are eventually validated or falsificated, i.e., some of the transactions made by malicious node violates the termination property, i.e., they cannot reach consensus to be falsificated. However, the affect to the liveness is very little since the invalid transactions have no impact on the functionality of this system, which is based on the validated transactions. Then, the validated transactions can be proved to be reliable and valid, which will be shown in the following subsection. However, unidentified invalid transactions could cause another problem, spamming, which will be addressed in Subsection~\ref{ss:reli}. 

% \iftoggle{disc}{}{We proposed two liveness condition: Liveness and Weak Liveness in Subsection~\ref{ss:val}. The Liveness condition is in general not feasible in our system since it is impossible to guarantee the validation of the transactions made by malicious nodes. Since there is no consensus on the individual transactions, there is no guarantee that the validity, or even the existence, of a transaction can be checked by other nodes, i.e., a transaction might never be known by other nodes if the related parties refuse to reveal it to others. However, this scenario has minimum affect on our system since this transaction will simply be ``ignored'' by all honest nodes. Also, it is not a necessity for implicit consensus.}

% It can be easily checked that the Weak Liveness condition is sufficient for implicit consensus. However, the Weak Liveness condition will fall into the pitfall of the FLP impossibility \cite{flp}, which proves that deterministic approach cannot deal with crash failures (offline node). For example, if a transaction $tr( i \to j,s)$ is validated by $i$. However, nodes $i,j$ all go offline immediately after the validation. This will result in the violation of the Weak Liveness condition.
% Here, we show that this problem also has minimum effect on the correctness of our system because of three reasons.

\subsection{Correctness}

%Since our validation process is deterministic, the correctness of our validation scheme can be proved if we show that the correct and complete proofs of a transaction are consistent and the result of the validation process with these proofs is correct. 
%Since different nodes might have different observation of the network, we firstly introduce the concept of a {\em view}.
Correctness condition guarantees the validity of our validation scheme, i.e., the validation result of the honest nodes will be consistent with the validity of the transactions, which is a global and unambiguous property of the transaction. 

Firstly, note that in our system there does not exist a globally agreed set of blockchains ${\cal B}_i, i\in\{1,2,\ldots,N\}$, i.e., in different time, nodes might have different observations of the blockchain set ${\cal B}_i$ due to latency or intended forking by malicious nodes. However, all the versions obtained by the honest nodes must be aligned with the already established consensus ${\cal CON}(r)$. Hence, we define the view of the blockchains in round $r$ as follows.

\begin{definition}[View]
A {\bf view} in consensus round $r$ denoted by $I(r)$ is a set of blockchains ${\cal B}_i, i\in{1,2,\ldots, N}$ with ${\cal CON}(r)$ as its consensus results.
\end{definition}

%Note that the view $I(r)$ could be different for different node at each round $r$. It could even be different for the same node at different time within a round. However, in each round, all possible views $I(r)$ will share the same consensus result ${\cal CON}(r)$ at any time for any honest nodes.
Basically, a view is the observation of the network by the honest nodes.
We now show that the position, order, and the content of the CPs are identical in all possible $I(r)$.

\begin{lemma}[Consistency of the CPs]\label{lm:fixcp}
If $B_i(k)$ and $B_i(\ell)$ are two blocks in the view $I(r)$, both of them are CPs included in the established consensus ${\cal CON}(r)$, and $B_i(k)$ is the previous CP of $B_i(\ell)$, then $B_i(k)$ is also the previous CP of $B_i(\ell)$ in any other view $I'(r)$. Moreover, $B_i(k)$ and $B_i(\ell)$ are identical to their counterpart in other views, respectively.
\end{lemma}

%Note that the fraudful transaction is not equivalent to an invalid transaction (a transaction which is not valid). For example, a transaction have been written in a piece in one node and have not yet been written in another piece in the other node. It is invalid but not fraudful, since it might become valid while the transaction message been written in a piece of blockchain of that node. Also, note that the valid (fraudful) transactions are not equivalent to validated (falsificated) transactions. The former is a global property and the latter depends on whether it passes the validation scheme run by an individual node. However, we will now prove that in our system, the result of the locally executed validation scheme (validated or falsificated) will also reveal its global property (valid or fraudful).

Then we will show that the CPs protect the consistency of the pieces of the chains, i.e., there cannot exist two distinctive pieces which start from the same CP or end by the same CP which are both correct.

\begin{lemma}[Consistency of the Pieces]\label{lm:auth}
If a piece of blockchain ${\cal B}=\{B_i(k),B_i(k+1), \ldots, B_i(\ell)\}$ in a view $I(r)$ is correct, then there does not exist another piece ${\cal B}'=\{B_i(k),B_i(k+1), \ldots, B_i(\ell')\}$ or ${\cal B}'=\{B_i(k'),B_i(k'+1), \ldots , B_i(\ell)\}$ in any view $I'(r'), r' \geq r$ that is correct.
\end{lemma}

By Lemma~\ref{lm:auth}, since the proofs of a transactions are simply a collection of pieces, we directly have the following theorem.

\begin{theorem}[Consistency of the Proofs]\label{cor:proof}
If ${\cal P}$ is the correct and complete proofs of a transaction $tr(i \to j,s)$ in a view $I(r)$, then there does not exist proofs ${\cal P}'\ne {\cal P}$ of the transaction $tr(i \to j,s)$ which are also complete and correct in any view $I'(r'), r' \geq r$.
\end{theorem}

%\subsubsection{Theorem}

With the established lemmas and theorems, we prove the main theorem for the correctness of the validation scheme.

\begin{theorem}[Correctness of the Validation Scheme]\label{th:corr}
Assume that $u$ is an honest node. Then, if $V_u(tr(i \to j, s)) = \mbox{\em validated}$, then $tr(i \to j, s)$ is valid. If $V_u(tr(i \to j, s)) = \mbox{\em falsificated}$, then $tr(i \to j, s)$ is invalid.
\end{theorem}

\section{Performance}\label{s:adv}

In this section, we compare the performance of our system to other blockchain systems with the focus on throughput and reliability.

\subsection{Throughput}

By design, the throughput of our system is independent of the consensus scheme since each node can create as many transactions as they could with no guarantee on validity. A fair throughput comparison should be between the rate of valid transactions in our system, i.e., the amount of total valid transactions made in our system per second, to the transactions rate of the other blockchain systems.
The valid transaction rate is determined by the validation process, which is then determined by the amount of proofs that are required for each transaction. At first glance, the amount seems to be a lot since the proofs do not only contain all the related transactions, but also the chains of the nodes who make those transactions. However, the collection is incremental and we will show that the throughput is actually at least as good as some of the existing techniques.

First of all, if the transaction pattern of the network is isolated, e.g., exists subgroups of the network that only have transactions within the subgroup. In that case, the proofs of such transactions will only contain chains of the nodes in that subgroup. As a result, our scheme has the same advantage as sharding \cite{ethshard,omni} and achieves unbounded performance. Since specific sharding mechanism is not needed and the nodes simply optimize the storage and transmission costs spontaneously, we call it spontaneous sharding. A detailed analyze is shown in Appendix~\ref{a:3}. The same scenario holds for the micro transaction scenario allowed by lightning network. When two nodes transact many transactions with each other, the proofs are no more than the chain of each other. This is no more complicated than the validation in lightning network. However, note that our scheme achieves an uncompromised reliability. Hence, unlike lightning network which only functions for micro transactions, the application of our scheme is not economically restricted.

On the other hand, if the transaction pattern is more correlated in the sense that there is no isolated subgroup, our scheme still has less message complexity than $O(N)$ since not all transactions are required to be collected by each node. More specifically, node $i$ do not need to store the chain of node $j$, if node $j$ has never made a transaction that is used as the source (or recursively sources of the sources) of node $i$. As a result, the performance of our scheme is strictly unbounded except for the very extreme transaction patterns. Moreover, rational nodes will try to avoid that situation by choosing sources that minimize the amount of information to transmit. In other words, if the network is rational, it tends to be sharded with our scheme.

%Since this throughput is significantly different from the {\em scalable} throughput achieved and stated in other works \cite{croman,scp,miller}, we claim that an ``unbounded'' throughput is achieved by our system,  In most literature, the term ``scalable'' is used to compare with the ``unscalable'' throughput of classical BFT algorithms, which have communication costs of at least $O(N^2)$ per transaction. Then, a scalable blockchain usually suggests a blockchain with a consensus algorithm with a communication cost of $O(N)$. Hence, the transaction rate of a scalable blockchain system is upper bounded by $O(C)$ even if the transactions are very uniformly distributed. So far as we know, our scheme is the first and only blockchain system which achieves unbounded performance, i.e., the transaction rate in between $O(C)$ and $O(CN)$.

Note that although our throughput is unbounded, the latency still depends on the BFT algorithm, thus not scalable. More precisely, the consensus is reached on the CMs with a size of $O(N)$. As a result, the latency would be high in a large network. However, we can reduce the latency by using more scalable and efficient BFT schemes like \cite{700bft,hybrid,miller} since our scheme is not restricted to a specific BFT algorithm.

\subsection{Reliability}\label{ss:reli}

%In many existing blockchain systems, the throughput improvement is achieved by sacrificing either the decentralization \cite{hyperledger,corda} or the reliability \cite{ng,stellar\iftoggle{disc}{}{, tangle}}.
%In our system, as discussed in Subsection~\ref{ss:ic}, going from explicit consensus to implicit consensus does not compromise on either the decentralization or the reliability of the network. Firstly, it is clear that our system is completely decentralized. Then, it has been shown that a transaction is validated by an honest node, it is as reliable as a valid transactions in classical blockchain systems with standard reliability assumptions. That is to say, a malicious node cannot convince an honest node that an invalid transaction is valid unless it controls more than $\lfloor N/3 \rfloor$ nodes and/or it breaks the hash function or the asymmetric encryption.

In has been proved that the reliability of validated transaction is the same as the traditional BFT schemes since the correctness and agreement hold for all honest nodes.
Certainly, as discussed in Subsection~\ref{ss:live}, the price we pay is the termination, that is, some of the invalid transactions made by malicious node cannot be falsificated. The undecided transactions themselves do very little harm to the reliability since honest nodes will not use undecided transactions as sources thus this ambiguity will not propagate. However, it does give rooms to the malicious nodes to spam invalid transactions to overwhelm the honest nodes. This problem is similar to the DDoS (Distributed Denial-of-Service) attack which can be solved by some reputation/blacklist scheme. Actually, we believe that keeping the record of the invalid transactions is beneficial to the reliability of the system, since it provides the necessary information for honest nodes to identify malicious nodes and take actions.

Another problem is the degradation in the reliability if we loosen the constraint of the honest nodes in Definition~\ref{def:honest} and allow honest nodes to be offline. This will harm the liveness condition since there is a chance that the proofs of valid transactions cannot be obtained when the nodes which have the proofs of this transaction all go offline.
However, this problem is actually solved by the logic behind our system and the ``self interest'' phenomenon, i.e., every node is responsible for its own transaction. In our system, a transaction is only valid if it is validated by other nodes. Hence, it is in the interest of at least one of the related parties to prove it to the other nodes. Furthermore, if a node wants to use a transaction as the source for its transaction, it not only needs to prove the validation of this transaction, but also needs to keep the proofs and show the proofs to the other related party.
The validation scheme is also censorship-free, which suggests that any node that has validated a transaction can independently show the complete and correct proofs to other honest nodes for the validation. In other words, malicious nodes cannot prevent the termination property of valid transactions. As a result, the valid transactions are as reliable as the traditional BFT schemes since the agreement and correctness properties hold and the termination property is guaranteed by the self-interest of related parties.

\section{Conclusions and Discussion}\label{s:fw}
In this paper, we proposed a value-exchange blockchain system with a novel consensus model, namely implicit consensus. Our system achieves significant improvements in throughput and other important aspects comparing to other blockchains techniques. We hope that the following proposed concepts would shed a light on the future blockchain research.
\begin{itemize}
\item Termination condition of the BFT is not mandatory in a value exchange system such as a distributed ledger. It will always be the interest of some nodes to prove the validity and agreement of the transactions to the rest of the network. Hence, we do not need extra mechanism to force that. Scalability could then be achieved by leaving the termination condition aside.
\item Side-chain transactions can be as reliable as they are on the main chain as long as the whole side-chains of all the sources are examined. This is not necessarily a heavy task since the amount of transactions that need to be validated is still only a fraction of the whole transactions set.
\item In our system, rational nodes will try to optimize their storage and message transmission by only validate and keep the proofs for their own transactions, which is a spontaneous sharding. Moreover, if multiple sources are available for a transaction, they will choose the one that requires minimum data transmission. Then, the sharding is also self-optimized.
\end{itemize}

\bibliography{Implicit_Consensus}

\begin{thebibliography}{10}

\bibitem{nakamoto}
Nakamoto, S.:
\newblock Bitcoin: A peer-to-peer electronic cash system.
\newblock (2008)

\bibitem{lamport}
Lamport, L., Shostak, R., Pease, M.:
\newblock The byzantine generals problem.
\newblock ACM Transactions on Programming Languages and Systems (TOPLAS)
  \textbf{4}(3) (1982)  382--401

\bibitem{flp}
Fischer, M.J., Lynch, N.A., Paterson, M.S.:
\newblock Impossibility of distributed consensus with one faulty process.
\newblock J. ACM \textbf{32}(2) (April 1985)  374--382

\bibitem{bracha}
Bracha, G.:
\newblock Asynchronous byzantine agreement protocols.
\newblock Information and Computation \textbf{75}(2) (1987)  130--143

\bibitem{castro}
Castro, M., Liskov, B.,  et~al.:
\newblock Practical byzantine fault tolerance.
\newblock In: OSDI. Volume~99. (1999)  173--186

\bibitem{elastico}
Luu, L., Narayanan, V., Zheng, C., Baweja, K., Gilbert, S., Saxena, P.:
\newblock A secure sharding protocol for open blockchains.
\newblock In: Proceedings of the 2016 ACM SIGSAC Conference on Computer and
  Communications Security. CCS '16, New York, NY, USA, ACM (2016)  17--30

\bibitem{700bft}
Guerraoui, R., Kne{\v{z}}evi{\'c}, N., Qu{\'e}ma, V., Vukoli{\'c}, M.:
\newblock The next 700 {BFT} protocols.
\newblock In: Proceedings of the 5th European conference on Computer systems,
  ACM (2010)  363--376

\bibitem{byzcoin}
Kokoris{-}Kogias, E., Jovanovic, P., Gailly, N., Khoffi, I., Gasser, L., Ford,
  B.:
\newblock Enhancing bitcoin security and performance with strong consistency
  via collective signing.
\newblock CoRR \textbf{abs/1602.06997} (2016)

\bibitem{croman}
Croman, K., Decker, C., Eyal, I., Gencer, A.E., Juels, A., Kosba, A., Miller,
  A., Saxena, P., Shi, E., Sirer, E.G.,  et~al.:
\newblock On scaling decentralized blockchains.
\newblock In: International Conference on Financial Cryptography and Data
  Security, Springer (2016)  106--125

\bibitem{ng}
Eyal, I., Gencer, A.E., Sirer, E.G., Van~Renesse, R.:
\newblock Bitcoin-{NG}: A scalable blockchain protocol.
\newblock In: 13th USENIX Symposium on Networked Systems Design and
  Implementation (NSDI 16), USENIX Association (2016)  45--59

\bibitem{ln}
Poon, J., Dryja, T.:
\newblock The bitcoin lightning network: Scalable off-chain instant payments.
\newblock Technical Report (draft) (2015)

\bibitem{algorand}
Micali, S.:
\newblock {ALGORAND:} the efficient and democratic ledger.
\newblock CoRR \textbf{abs/1607.01341} (2016)

\bibitem{hybrid}
Pass, R., Shi, E.:
\newblock Hybrid consensus: Efficient consensus in the permissionless model
  (2016)

\bibitem{sleepy}
Bentov, I., Pass, R., Shi, E.:
\newblock The sleepy model of consensus.
\newblock IACR Cryptology ePrint Archive \textbf{2016} (2016)  918

\bibitem{ethshard}
Buterin, V.:
\newblock On sharding blockchains.
\newblock Sharding FAQ (2017) Available at
  \url{https://github.com/ethereum/wiki/wiki/Sharding-FAQ}.

\bibitem{omni}
Kokoris-Kogias, E., Jovanovic, P., Gasser, L., Gailly, N., Ford, B.:
\newblock Omniledger: A secure, scale-out, decentralized ledger.

\bibitem{scp}
Luu, L., Narayanan, V., Baweja, K., Zheng, C., Gilbert, S., Saxena, P.:
\newblock {SCP}: A computationally-scalable byzantine consensus protocol for
  blockchains.
\newblock IACR Cryptology ePrint Archive \textbf{2015} (2015)  1168

\bibitem{miller}
Miller, A., Xia, Y., Croman, K., Shi, E., Song, D.:
\newblock The honey badger of {BFT} protocols.
\newblock In: Proceedings of the 2016 ACM SIGSAC Conference on Computer and
  Communications Security, ACM (2016)  31--42

\end{thebibliography}

\newpage

\appendix

\section{Proofs}

\subsection{Proof of Theorem~\ref{th:pl}}
\begin{proof}
By the definition of honest nodes, $i$ and $j$ will add the transaction messages to their chains. The messages will be included in a correct piece before some time $t$ because of the conditions of the consensus scheme. Then, the correct and complete proofs of this transactions can be obtained by honest nodes since $i$ and $j$ are honest, which suggests that the outcome of the validation scheme will not be ``undecided'' and complete the proof.
\end{proof}

\subsection{Proof of Lemma~\ref{lm:fixcp}}
\begin{proof}
The proof follows from the definition of the CM and the consensus scheme. By the definition of the CM, the information of the position, order, and the digests of the content of the CPs are included in the CMs. Moreover, the CMs with incorrect information or the ones that attempts to create forks in CPs are discarded during the consensus process. As a result, the consensus ${\cal CON}(r)$ fixes the position, order, and the content of the CPs. Then, this lemma is established if the two views $I(r)$ and $I'(r)$ have the same consensus results.
\end{proof}

\subsection{Proof of Lemma~\ref{lm:auth}}
\begin{proof}
We prove this lemma by contradiction.

Assume there exists another correct piece of blockchain ${\cal B}'=\{B_i(k'), B_i(k'+1), \ldots, B_i(\ell') \},$ $ k'\ne k$ or $\ell' \ne \ell$ that ${\cal B}' \ne {\cal B}$ in a view $I'(r')$. By the definitions of a correct piece, we know that $B_i(k), B_i(k'), B_i(\ell), B_i(\ell')$ are all CPs included in ${\cal CON}(r)$. Moreover, by Lemma~\ref{lm:fixcp}, we have $\ell=\ell', B_i(\ell)=B_i(\ell')$ if $k=k'$ and $k=k', B_i(k)=B_i(k')$ if $\ell =\ell'$. 

Then, since both ${\cal B}$ and ${\cal B}'$ are correct, all digests of all blocks in these pieces should be correct. Then, since ${\cal B} \ne {\cal B}'$, there must exists two blocks $B_i(n)\ne B'_i(n)$ such that $B_i(n+1)=B'_i(n+1)$, which suggests $H(B_i(n))=H(B'_i(n))$. This contradicts the fact that the digests are collision free.
\end{proof}

\subsection{Proof of Theorem~\ref{th:corr}}
\begin{proof}
We firstly proof the following statement: If $V_u(tr(i \to j, s)) = \mbox{\em validated}$ and all of its sources are valid, then $tr(i \to j, s)$ is valid. If $V_u(tr(i \to j, s)) = \mbox{\em falsificated}$ and none of its source are falsificated, then $tr(i \to j, s)$ is invalid.
We prove this by contradiction. 

Firstly, assume that there exist an invalid transaction $tr(i \to j, s)$ with valid sources and it is validated by an honest node $u$. Then, the correct and complete proofs of this transaction must have been collected by $u$. Furthermore, it must have passed the validation process. Then, since the steps in validation process (Definition~\ref{def:vp}) are precisely the validity conditions (Definition~\ref{def:vali}) except the {\bf Correct Chains}, which has already been guaranteed by the proof collection. By Definition~\ref{def:vali}, there exists an observation of this network in which all the validity conditions for this transactions are met, thus this transaction is valid. This contradict our assumption.

Then, we assume that there exists a valid transaction $tr(i \to j, s)$ with no falsificated source and it is falsificated by an honest node $u$. By the definition of the validation scheme, node $u$ must have collected all the proofs for this transaction, which includes all the proofs for the sources. Hence all of its source are validated, thus are valid. Then, at least one of the items in the Definition~\ref{def:vp} except the {\bf Validated Sources} is violated, which suggests that the validity conditions (Definition~\ref{def:vali}) are not fulfilled. Then, by Theorem~\ref{cor:proof}, the proofs are consistent. Hence, there does not exist an observation by an honest node in which all conditions are satisfied. By Definition~\ref{def:vali}, this transaction is invalid, which contradicts our assumption.

The theorem is thus proved by recursively using the proved statement on the transactions and their sources since the validity of the initial balance can be checked with ${\cal CON}(r)$. 
\end{proof}

\section{Analysis of the Throughput}\label{a:3}
Here we lower bound the rate of the valid transactions in our system. For the sake of easier comprehension, we assume that the transactions rate, communication capacity, and computation capacity are uniform for all nodes and all time and the adversaries do not spam invalid transactions.

%Intuitively, a best case scenario would be that nodes only do transactions with one other node. In that case, the validation of transactions by one node only requires the chain of the other node as proofs and validates all the transactions in that chain. Also, for the other node, it only needs to provide its chain to one node only.
% %As a result, we have the individual validation rate equals to $O(C)$, where $C$ is the validation capacity of that node. Then, we have the throughput of our system $O(NC)$ since the nodes can validate different transactions in parallel.
% %It is a remarkable and even counter intuitive result since it suggests that the throughput actually grows proportionally as the network size grows. However, it does make sense in a setting as suggested above, where basically nodes only do transactions with one other node and do not mind or bother to validate any other transactions. Here, we show that for any kind of network partition, the throughput is always in between $O(N)$ and $O(1)$. For the sake of easier comprehension, we assume that the transactions rate, communication capacity, and computation capacity are uniform for all nodes and all time and the adversaries does not spam invalid transactions.

We consider a subset of node in the network ${\cal G}, |{\cal G}|=g\leq N$ which only do transactions with the nodes in the subset. Assume that each chain grows with a rate of $R$ messages/second and the duration of a consensus round is $T$. The amount of messages generated by this subset of nodes in a round is $RgT$, which can be divided into two parts: valid transactions and invalid transactions. Since the honest nodes only make transactions that they can validate, the amount of valid transactions is at least $R_vgT$ where $R_v$ is the validation rate. The invalid transaction can only be made by adversaries. Since they do not spam, we have the amount of the invalid transactions equals to $R_agT$ where $R_a = O(R_v)$. Then, we have $R=R_v+R_a=O(R_v)$. %Since the adversaries do not spam invalid transactions, 
% %we have
% %\begin{equation}\label{eq:rv}
% %$R_v=\alpha R$,
% %\end{equation}
% %in which $R_v$ is the rate of validated transactions of each node and $\alpha \in (0,1]$ is a constant that only depends on the behavior of the adversaries and the ratio of the adversaries in the network.

Let us analyze the duration that a node needs to validate all transactions that it makes in a round.
For the proof collection, it needs no more than all chains in ${\cal G}$, which requires data transmissions with no more than an amount of $RgT$ messages since the proof collections is incremental, i.e., only the newly generated parts of the chains are needed. The proofs are collected based on point-to-point transmissions. Each node broadcasts its chain at a rate of $C_{comm}/g$, where $C_{comm}$ is the communication capacity (message/second) of the nodes. The collection rate is then $C_{comm}$ since nodes broadcast their chains simultaneously in different channels. Hence, we have the duration of proof collection
% %\begin{equation}\label{eq:pc}
$t_p \leq \frac{RgT}{C_{comm}}$.
% %\end{equation}

For validation, in the worst case, all of these transactions need to be validated, which requires duration
% %\begin{equation}\label{eq:vp}
$t_v \leq \frac{RgT}{C_{comp}}$,
% %\end{equation}
where $C_{comp}$ is the computation capacity (message/second).
By basic queuing theory, we should have
% %\begin{equation}\label{eq:queue}
$t_p+t_v = T$.
% %\end{equation}

Then, since honest nodes only make transactions that they can validate and the in all the $RgT$ messages, the expected invalid message
Since all validated transactions are valid (Theorem~\ref{th:corr}), combining all the inequalities above, we have a lower bound on the rate of the valid transactions for each node
% %\begin{equation}
$R_v \geq \Omega(\frac{C}{g})$,
% %\end{equation}
where $C=\frac{C_{comm}C_{comp}}{C_{comm}+C_{comp}}$. Then, the throughput of this group is lower bounded by $\Omega(C)$ since a group has $g$ nodes that can simultaneously make transactions.

This lower bound suggests that the throughput in any separate group of nodes in the network is completely independent of the rest of the network and only depends on the communication and computation capacity of the nodes in that group. This is an ideal property to have for a blockchain system since the throughput is no longer limited by the throughput of the consensus algorithm. In the best case that nodes are paired and only do transaction with each other, we achieve a throughput of $O(CN)$. In the worst case that all nodes make transactions with all other nodes, we achieve a throughput of $O(C)$.
 \end{document}